\documentclass[sigconf]{acmart}
\usepackage[]{graphicx}\usepackage[]{color}
\makeatletter
\def\maxwidth{ %
  \ifdim\Gin@nat@width>\linewidth
    \linewidth
  \else
    \Gin@nat@width
  \fi
}
\makeatother

\definecolor{fgcolor}{rgb}{0.345, 0.345, 0.345}

\usepackage{framed}
\makeatletter
 {\par\unskip\endMakeFramed%
 \at@end@of@kframe}
\makeatother

\definecolor{shadecolor}{rgb}{.97, .97, .97}
\definecolor{messagecolor}{rgb}{0, 0, 0}
\definecolor{warningcolor}{rgb}{1, 0, 1}
\definecolor{errorcolor}{rgb}{1, 0, 0}

\usepackage{alltt}

 \usepackage{multirow}

\newcommand{\be}{\begin{equs}}
\newcommand{\ee}{\end{equs}}

\DeclareFontFamily{U}{matha}{\hyphenchar\font45}
\DeclareFontShape{U}{matha}{m}{n}{
  <-6> matha5 <6-7> matha6 <7-8> matha7
  <8-9> matha8 <9-10> matha9
  <10-12> matha10 <12-> matha12
  }{}

\DeclareSymbolFont{matha}{U}{matha}{m}{n}
\DeclareMathSymbol{\Lt}{3}{matha}{"CE}

\graphicspath{{./}{input/}}

\makeatletter
\providecommand*{\input@path}{}
\g@addto@macro\input@path{{./}{input/}}
\makeatother

\acmConference[FAT/ML 2017]{4th Workshop on Fairness, Accountability, and Transparency in Machine Learning}{August 2017}{Halifax, Nova Scotia, Canada} 
\acmYear{2017}
\IfFileExists{upquote.sty}{\usepackage{upquote}}{}
\begin{document}

\title{The causal impact of bail on case outcomes for indigent defendants} 

\author{Kristian Lum}
\affiliation{
\institution{Human Rights Data Analysis Group}
\city{San Francisco}
\state{CA}
\postcode{94110}
\country{USA}
}

\author{Mike Baiocchi}
\affiliation{%
  \institution{Stanford University}
  \department{Stanford Prevention Research Center}
  \city{Palo Alto}
  \state{CA}
  \postcode{94305}
  \country{USA}
}

\maketitle

\renewcommand{\shortauthors}{Lum and Baiocchi}

\begin{abstract}
We use near-far matching, a technique for estimating causal relationships, to explore whether bail causes a higher likelihood of conviction. 
\end{abstract}

\keywords{ pre-trial detention, money bail, bail, observational studies, causal inference, near-far matching, matching}

\section{Introduction} 

In the United States, the money bail system has come under recent scrutiny due to its contribution to mass incarceration and its impact on poor defendants \cite{Watch:aa}.  Under the current system, a judge may choose to set an amount of money (bail) that is required in order to secure the accused's release from detention prior to standing trial. If the defendant can pay the bail-- either from personal or familial funds or through a bail bondsman -- the defendant is released from detention. If the accused cannot raise the funds, he must remain incarcerated until the case is resolved, whether by plea or trial. 

It has long been observed that those who are detained pre-trial are more likely to be convicted \cite{ares1963manhattan,rankin1964effect, phillips2007pretrial, phillips2008pretrial, warren1972unconstitutional}, but only recently have formal causal inference methods been brought to bear on the problem of determining whether pre-trial detention {\it causes} a higher likelihood of conviction \citep{gupta2016heavy, leslie2016unintended, stevenson2016distortion, dobbie2016effects}. In each case where causal inference methods were used, a statistically significant effect was found.
	
In this paper, we also apply statistical methods to the problem of quantifying the impact of setting bail on case outcome. Unlike the previous mentioned studies, ours focuses specifically on the population of defendants represented by an organization of public defenders, making this a particularly vulnerable population deserving of specific attention. Our work is further differentiated from previous work in this area in that we approach the problem using methodology from a different tradition for estimating causal effects. Whereas previous studies have employed a stwo-stage model-based instrumental variable approach that is common in econometrics, we approach this problem using near-far matching \cite{baiocchi2010building, baiocchi2014instrumental}, a matching procedure that derives from the ``observational studies" tradition. To our knowledge this is the first use of matching techniques in this specific setting.   

The work described here is a bit of a departure from that typically presented at FAT ML in that we are not developing a new machine learning algorithm with fair outcomes in mind nor are we critiquing existing algorithms. Instead, what follows is more along the lines of an applied statistics project that, at its core, is concerned with fairness and accountability in the real world. That is, we assess the fairness of the current money bail system in terms of its potential to cause those who cannot afford bail to be more likely to plead or be found guilty. Due to length limitations put in place by the FAT ML conference, figures, tables, and discussion that would normally appear to support a causal analysis have been omitted in this preliminary work.

\section{Data} \label{sec:data}
\subsection*{i. Inclusion/exclusion}
Our dataset consists of all felony and misdemeanor cases that were handled from start to finish by an attorney from our partner public defender agency in 2015. We do not consider cases in which the defendant was extradited, the case was transferred to a special court (e.g. family court), or very irregular cases (e.g. the crime was abated by the death of the client). We also do not consider cases that were disposed at arraignment, i.e. cases in which the defendant immediately plead guilty or the judge dismissed the charges at arraignment.  We do not consider these cases to be part of the population of interest because in these cases bail cannot be set and the concept of pre-trial detention is irrelevant, as there is to be no further trial and there is no pre-trial period of which to speak.

\subsection*{ii. Covariates}
Our dataset includes a variety of demographic covariates about the individual -- age at the time of the alleged crime, gender, race, and ethnicity. At the time of processing in the intake interview, the defendant is also asked to report their employer, weekly income, phone number, and address. We include an indicator of whether the defendant declared an employer, their self-reported weekly income, an indicator for whether they reported a phone number, and an indicator of whether they reported an address. Although this is all self-reported information, this is the same information that is available at the time of arraignment, and thus is the information relevant to determining whether bail will be set. Last, as a measure of prior criminal activity, we include the number of prior counts for which the public defender's office represented that client in the previous year (2014). This is a noisy measure of prior criminal activity, as it is possible that in 2014, the defendant had additional charges but had different legal representation. It is important to note that, despite the fact that there may be relevant variables (like a longer criminal history variable) that are omitted from the analysis, the instrumental variable method we employ allows us to obtain causal estimates nonetheless. 

For each case, our dataset also includes information about the charges against the defendant-- the type of the offense (misdemeanor or felony), the most serious charge against the defendant (the ``top charge"), the class of the most serious charge (A, B, C, D, E).
The top charge in the case is a  specific category that describes both the nature of the crime as well as the severity, in most cases denoted by the number following the crime description, e.g. ``Assault 3".

\subsection*{iii. Outcome and treatment}
\subsubsection*{Outcome}
The outcome variable of interest, $G$, is an indicator of whether the defendant was found guilty. Specifically, we set $G=1$ if the final disposition of the case was a guilty plea (the defendant plead guilty without going to trial) or a guilty verdict (the case went to trial, and the defendant was found guilty), and $G=0$ if the case was dismissed (the charges were dismissed without going to trial) or the defendant was acquitted (the case went to trial, and the defendant was found not guilty). We treat $G$ as missing if a final determination has yet to be made.

\subsubsection*{Treatment}
 As the treatment variable, $T$, we use an indicator that denotes whether bail was set in the case. Although our hypothesis for how the money bail system causes worse case outcomes revolves around pre-trial detention-- not the setting of bail-- pre-trial detention only occurs if bail is set and the defendant cannot pay. Thus the proximate cause of pre-trial detention is the setting of bail. 

\subsection*{iv. Instrumental variable} 
For the instrumental variable (IV), we calculate a measure of judge strictness or severity. To conform to the conventions of near-far matching, we calculate this so that low levels of the IV correspond to more strict judges, and higher levels of the IV to more lenient judges. Several other analyses have used judge severity as a pseudo-randomizer \cite{martin1993special, aizer2015juvenile, kling2006incarceration}. In particular, \cite{kleinberg2016human} use a similar instrumental variable in an analysis how judges  determine to whom they grant pre-trial release and the likelihood with which individuals who would have been released would have failed to appear in court. Other studies also rely on judge randomization or quasi-randomization in assessing the causal impact of incarceration or probation on recidivism \cite{green2010using, berube2007effects}. 

At the core of all of these analyses is the assumption that some part of the decision-making process depends on features of the judge, rather than the facts of the case, and that defendants are pseudo-randomized to judges. Though we do not have room to describe the process by which defendants are assigned an arraignment judge here, we believe that this procedure meets the standards for pseudo-randomization.  Our identification strategy makes use of the insight that some judges are predisposed to set bail (``strict") and others are less likely (``lenient"). More technically, our ``pseudo-randomizer" is a judge's rate of granting pre-trial release without bail, for a specific crime type, relative to other judges in that region.  

We calculate the instrumental variable separately for each administrative region and crime (e.g. Assault 3), resulting in a judge-region-crime-specific measure of severity. Similar to \citep{gupta2016heavy,leslie2016unintended}, we use a leave-one-out method for calculating this variable so that the $i$th defendant's own outcome does not influence the calculation of the instrumental variable for his case. Let $T_{jbci}$ denote the treatment variable (1 if bail was set; no otherwise) of the $i$th individual seen by judge $j$ in region $b$ with top charge $c$. We define $T'_{jbci} = 1-T_{jbci}$.  Then, we calculate judge severity measure for the $i^*$th defendant as follows:

\begin{equation}
S_{jbc}^{(i^*)} = \frac{1}{n_{jbc}-1}(\sum_i T'_{jbci}  - T'_{jbci^*})  - \frac{1}{n_{bc}-1}  (\sum_{i, j} T'_{jbci} - T'_{jbci^*}),
\end{equation} 

\noindent where $n_{jbc}$ is the number of cases with top charge $c$ seen by judge $j$ in region $b$, and $n_{bc}$ is the number of cases with top charge $c$ seen in region $b$. 

\section{Method} \label{sec:method}
 
Current recommendations for best practices in observational studies of medical interventions typically favor a matching approach, rather than the two-stage model-based inference that is popular in econometrics \cite{pcori2013pcori}. In this study, we use near-far matching. The logic of near-far matching follows the design of a randomized experiment that suffers from noncompliance with the randomization - this is sometimes called an ``encouragement design" \cite{holland1988causal}. In encouragement randomized trials, some physicians are randomly assigned to be ``encouraged" to perform or suggest a particular treatment to their patients, others are not \cite{dexter1998effectiveness}. The result is that some patients, even after accounting for their own personal attributes or the severity of their condition, are more likely to receive the treatment due only to the level of ``encouragement" their physician  received. Analogous observational studies in which there is patient-independent variability in the physician-specific inclination towards a treatment can be undertaken in this setting if patients are pseudo-randomized to physicians. Methodology for these studies exploits this randomized push towards receiving the treatment to isolate the ``natural experiment'' that exists in the data \cite{zubizarreta2014isolation}. 

Near-far matching mimics a randomized encouragement trial by preferentially creating matched pairs of observations that are (i) as nearly identical in pre-exposure variables as possible (``near in covariates"), while (ii) being as dissimilar as possible in their pseudo-randomized push to either be exposed or unexposed (``far in their encouragement"). Pairwise covariate proximity is measured by calculating the Mahalanobis distance between covariate vectors. A non-bipartite matching algorithm is then used to find a set of pairings that minimizes the Mahalanobis distance between the matched pairs while maximizing the pairwise difference in the instrumental variable.  In our case, this would look like finding two identical defendants -- that is, who looked the same in all ways measured in our data set prior to the bail-setting hearing, but one defendant was routed to a ``strict" judge and the other defendant was routed to a ``lenient" judge. Note that within this pair we are attempting to isolate the judge's predisposition and use it as the determining factor for bail-setting, rather than allowing differences in the facts-of-the-case being the determining factor. 
 
Matching-based study designs focus heavily on the task of identifying reasonable comparator groups and limiting the analysis to those observational units. That is, we exclude observations because the real world data set did not give rise to suitable comparators. Most modern matching algorithms have a optimal ways for finding the most ``dissimilar" or ``uninformative'' units and removing them from the analysis. In our study we use sinks-- ``phantom" observational units that have the unique property that they are perfect matches to all real data points. The matching algorithm then runs on the augmented data set-- real and ``phantom" observations. The algorithm will tend to pair hard-to-match observational units to the sinks. In this implementation of near-far matching, we automatically select the optimal number of sinks by maximizing the $F$-statistic of a hypothesis test that measures the ``strength" of the instrumental variable, i.e. the degree to which encouragement correlates with treatment assignment. If a real observation is matched to a sink then we remove that observational unit from our analysis. 

The output of the matching procedure is a set of matched pairs, $\{i_1, i_2\}$ for $i=1, ..., I$, where $i_1$ and $i_2$ are is the indices of the encouraged and unencouraged defendants, respectively, in the $i$th pair. Then, for example, $G_{i_1}$ and $G_{i_2}$ are the case outcomes for the $i$th matched pair. Similarly, $T_{i_1}$ and $T_{i_2}$ are the the indicators of whether bail was set for the encouraged and unencouraged defendants, respectively, in the $i$th matched pair. Although not immediately obvious, the instrumental variable is embedded in the subscript notation, as those defendants who had high values of the IV are assigned to $i_1$ and those with low levels of the IV to $i_2$. Having obtained matched pairs, inference is then a relatively straightforward. The causal relationship is measured by estimating the ``effect ratio" as,

$$\lambda = \frac{\sum_{i=1}^I G_{i_1} - G_{i_2}}{\sum_{i=1}^I T_{i_1} - T_{i_2}}.$$

This quantity deserves a bit of attention to aid interpretation. In technical terms, one can describe $\lambda$ as a complier average causal effect of the risk difference which is conditional on the matched set. It measures the ratio of the difference in outcome between the encouraged and unencouraged groups to the difference in treatment. More informally, one can describe the estimate as the increase in probability of conviction due to bail setting for those defendants whose bail determination was likely to switch based on the type of judge that presided over the arraignment. 

The specific details of our matching procedure are as follows. We first stratify all defendants in our dataset into top charge-region-gender groups. These are the groups within which pairs will be created, forcing an ``exact match" on the top charge, region, and defendant gender of the case. For example, female defendants in region A whose top charge was Criminal Mischief 2 will only be matched to other female defendants in region A whose top charge was Criminal Mischief 2. Within these groups,  we use the nearfar package in the R computing environment to match similar defendants to one another \cite{Rigdon:2016aa}. The output of this procedure is a set of pairs of same-gender defendants who are each accused of identical crimes in the same region and who are maximally similar on all other covariates. Paired defendants differ in that they were arraigned by judges with differing levels of severity. Not all defendants are paired-- some are dropped as described to achieve the best possible inference.

\section{Results} \label{sec:results}
\subsection*{Covariate balance}

After the matching procedure is complete, we are left with $n =$ 61,486 defendants in our study.  We first assess whether our matching algorithm has successfully achieved covariate balance between the two groups, i.e. whether the encouraged group is similar to the discouraged group in terms of its observable covariates.  This is shown in Table \ref{tab:std_diffs-FALSE}. We  find that we were able to obtain excellent balance. For all covariates (i.e. all variables except the treatment, IV, and outcome, which are not meant to be minimized), we attained a standardized difference of less than 0.01. That is, the average difference between the encouraged and unencouraged defendant in each pair for each covariate was less than 1\% of one standard deviation.  This far exceeds the accepted standard that the standardized differences ought to be less than 10\% \cite{silber2001multivariate}.  These tables do not include a charge or region variable because defendants were matched only to other defendants who shared the same top charge and region. So, in some sense, these tables under-state the degree of balance by not explicitly showing that we have attained perfect balance on top charge and region. 

\begin{table}[ht]
\centering
\begin{tabular}{rrrr}
  \hline
 & Encouraged & Unencouraged & St Dif \\ 
  \hline
Guilty & 0.41 & 0.40 & 0.03 \\ 
  Bail Set & 0.21 & 0.16 & 0.12 \\ 
  IV & -0.07 & 0.07 & 1.18 \\ 
  Age & 32.69 & 32.71 & 0.00 \\ 
  White & 0.28 & 0.28 & 0.00 \\ 
  Black & 0.52 & 0.52 & 0.00 \\ 
  Non-Hispanic & 0.65 & 0.65 & 0.00 \\ 
  Male & 0.81 & 0.81 & 0.00 \\ 
  Prior Records 2014 & 0.54 & 0.53 & 0.00 \\ 
  Wkly Income & 53.00 & 52.75 & 0.00 \\ 
  Any Income & 0.12 & 0.12 & 0.00 \\ 
  Employer & 0.17 & 0.17 & 0.00 \\ 
  Phone Number & 0.15 & 0.15 & 0.00 \\ 
  Address & 0.91 & 0.91 & 0.00 \\ 
   \hline
\end{tabular}
\caption{Table of post-match standardized differences. Summary of data analyzed.} 
\label{tab:std_diffs-FALSE}
\end{table}

\subsection*{Generalizability}
Because our methodology drops some participants from the study so that we can obtain optimal matching, the next question to address is whether our matched sample-- the population from which we will make estimates-- is informative about the full dataset-- the clients of our partner public defender in 2015. Though omitted for space, figures showing side-by-side comparisons of the population used in the analysis to the full study population for each covariate show no substantive difference in distribution. Thus we believe that the results from our matched group are generalizable to an analysis of the study population. 

\subsection*{Estimates}

Table \ref{tab:estimatesFALSE} shows our estimates of $\lambda$, our measure of the causal impact of setting bail on the outcome of the case. The Est column displays a point estimate of $\lambda$. The Lo and Hi columns give the end points of a 95\% confidence interval. The column labeled as $n$ reports the number of observations in each stratum. The final column indicates whether the estimates are statistically significant at the $\alpha=0.05$ level.  We focus attention on the estimate at the top of Table \ref{tab:estimatesFALSE} referred to as the total estimate in the aggregate stratum. This is the global estimate across all case and defendant types and the focus of this study. This estimate should be interpreted as follows: for every additional 100 defendants that are assigned bail simply because they saw a stricter judge, an additional 34 guilty pleas or convictions will result that otherwise would not have. This represents a contextually meaningful increase in the probability of a guilty finding if bail is set. 

We also present stratum-specific estimates for a variety of stratification schemes. The focus of our analysis is on the aggregate effect estimate, though we report the others for completeness. For many stratum-specific estimates, there is insufficient data to obtain estimates with small enough confidence intervals to definitively determine whether there was a positive or negative impact. In some cases, the absolute value of the end points of the confidence interval exceeds 1, resulting in an estimated interval that extends beyond the possible range for an estimate that corresponds to an increase in probability. Although the interval contains out-of-bounds values, we report them as-is to emphasize the instability of those particular estimates. Due to the reduced sample sizes from stratifying and resulting reduction in statistical power, statistically significant differences between stratum-specific estimates are not possible. However, these stratum-specific estimates are suggestive sub-analyses that can be used to guide future research.  

\begin{table}[ht]
\centering
\begin{tabular}{lllllll}
  \hline
 & Stratum & Est & Low & Hi & n & * \\ 
  \hline
\hline
Aggregate & total & 0.34 & 0.2 & 0.49 & 56734 & * \\ 
   \hline
\multirow{5}{*}{Region} & A & 0.43 & 0.23 & 0.63 & 17010 & * \\ 
   & B & 0.34 & 0.14 & 0.54 & 17936 & * \\ 
   & C & -0.07 & -0.52 & 0.33 & 7290 &  \\ 
   & D & 0.66 & 0.13 & 1.35 & 12174 & * \\ 
   & E & 0.88 & 0.11 & 2.89 & 2324 & * \\ 
   \hline
\multirow{2}{*}{Crime Type} & Felony & 0.22 & -0.12 & 0.58 & 8448 &  \\ 
   & Misd. & 0.37 & 0.22 & 0.53 & 48286 & * \\ 
   \hline
\multirow{2}{*}{Gender} & Male & 0.31 & 0.16 & 0.45 & 46118 & * \\ 
   & Female & 0.65 & 0.12 & 1.27 & 10532 & * \\ 
   \hline
\end{tabular}
\caption{Estimated causal impact of setting bail on judicial outcome} 
\label{tab:estimatesFALSE}
\end{table}

\subsection*{Sensitivity Analysis}

Inherent to any methodology that relies upon instrumental variables is an unverifiable assumption regarding the IV's relationship to the  (unobserved) covariates, treatment, and outcome.  If these assumptions are unmet, it is possible to estimate a causal relationship where none exists. We perform a sensitivity analysis to assess the robustness of our inference that there exists a positive causal relationship between setting bail and conviction in a case.

Using the method described in \cite{baiocchi2010building}, we find that in order for the inference that there exists a positive causal relationship to be false, it would have to be the case that there is some excluded variable that increases both one's odds of assignment to a strict judge and also one's odds of conviction substantially. The magnitude of this increase would have to be similar to increasing the odds of assignment to a strict judge by a third and increasing the odds of conviction by half. Because of our confidence in the psuedo-randomization process to judges, we believe that the  departure from randomization necessary to nullify our results is unlikely.

\section{Discussion} \label{sec:discussion}
We find a strong causal relationship between setting bail and the outcome of a case for the clients of our partner public defenders-- specifically, we find that for cases for which different judges could come to different decisions regarding whether bail should be set, setting bail results in a 34\% increase in the chances that they will be found guilty. Though we approach the problem using a different tradition for analyzing observational data than other similar studies, our substantive findings support the conclusions of the recent literature in this area. That our estimate is significantly higher than reported in other recent work is consistent with our hypothesis that the effect of setting bail is likely stronger among vulnerable populations, such as those who rely on public defenders. It is also likely that our estimate deviates from other, recently reported estimates because of how we define the population to which the estimates pertain. For example, one of the recent cited studies considered only felonies. And, though it is not explicitly mentioned, it seems that several studies include cases that were disposed at arraignment, whereas we define our population to be cases that have made it past that stage. Regardless, combined with the other recent studies on the causal impact of setting bail, our study adds to the mounting empirical evidence that bail causes worse case outcomes. The real world implications of this are that there are likely many people--disproportionately, poor people-- who have been convicted of crimes simply because bail was set.

\newpage

\bibliography{bellamy.bib}

\bibliographystyle{ACM-Reference-Format}

\end{document}